\documentclass{aip-cp}

\usepackage[numbers]{natbib}
\usepackage{rotating}
\usepackage{slashed}

\usepackage{graphicx}
\usepackage{dcolumn}
\usepackage{bm}
\usepackage{epsfig}

\def\lam{\lambda}

\def\call{{\cal L}}

\def\calv{{\cal V}}

\def\beq{\begin{equation}}
\def\ee{\end{equation}}
\def\eeq{\end{equation}}

\def\bfig{\begin{figure}}
\def\efig{\end{figure}}
\def\bea{\begin{eqnarray}}
\def\bwt{\begin{widetext}}
\def\ewt{\end{widetext}}
\def\beann{\begin{eqnarray*}}
\def\eea{\end{eqnarray}}
\def\eeann{\end{eqnarray*}}

\def\nn{\nonumber}

\def\3p0{$^{3}P_{0}$}

\def\hs{\hspace{.5cm};\hspace{.5cm}}

\begin{document}

\title{Massive dark photons in a Higgs portal   model  }

\author[aff1]{Dimiter Hadjimichef\corref{cor1}}

\affil[aff1]{Instituto de F\'{\i}sica, Universidade Federal do Rio Grande do  Sul, Brazil}
\corresp[cor1]{dimiter.hadjimichef@ufrgs.br}

\maketitle

\begin{abstract}

 An extension of the Standard Model  with a hidden sector which consists
of  gauge singlets (a Dirac fermion $\chi$  and a    scalar $S$) plus a   vector boson  $V_\mu$ (dark massive photon) is studied.
The singlet scalar interacts with the Standard Model sector through the triple and quartic scalar interactions,
while the singlet fermion and  vector boson field interact with the Standard Model  only via the singlet scalar. The scalar
field generates the vector boson's mass. Perspectives for future $e^{-}e^{+}$ colliders is considered. 
 
\end{abstract}

\section{INTRODUCTION}

The enigma of dark matter (DM) remains unsolved.
Still one of the most evasive and fascinating mysteries in physics,  the problem of the dark
matter in the Universe stirs the imagination of  most astronomers, cosmologists and particle physicists, 
that are convinced that at least 90\% of the mass of the Universe is due to some non-luminous form of matter.
Modernly, dark matter candidates converge to a
variety of interesting and plausible candidates namely
the {\it weakly-interacting massive particles} (WIMPs):  Standard Model neutrinos \cite{wimp1}, sterile neutrinos \cite{wimp2},
axions \cite{wimp3},  supersymmetric candidates [neutralinos, sneutrinos, gravitinos, axinos] \cite{wimp4}-\cite{wimp6}
 light scalar dark matter \cite{wimp7}, dark matter from little Higgs models \cite{wimp8}, 
Kaluza-Klein particles \cite{wimp9},
superheavy dark matter \cite{wimp10}. An excellent review in theoretical and experimental aspects of
dark matter can be found in \cite{wimp11}. In
general they are present in theories of weak-scale physics
beyond the Standard Model (SM) and give rise to appropriate relic abundance. Calculations have shown that
stable WIMPs can remain from the earliest moments of
the Universe in sufficient number to account for a significant fraction of relic dark matter density. This raises
the hope of detecting relic WIMPs directly by observing
their elastic scattering on targets. In the dark matter zoo
many different types of particles have been introduced
and their properties theoretically studied.

 In the present work  a renormalizable extension of the SM with a hidden sector which consists
of SM gauge singlets (a singlet Dirac fermion $\chi$ and a singlet  scalar  $S$) plus a  vector boson  $V_\mu$, which we shall call
the {\it dark massive photon}, is studied.
The singlet scalar interacts with the SM sector through the triple and quartic scalar interactions. 
The singlet fermion and  vector boson field interact with the SM  only via the singlet scalar. The scalar
field generates the vector boson's mass.

\section{THE MODEL}

The model  is defined by the following Lagrangian density

\bea
\mathcal{L}&=&\mathcal{L}_{SM}+\mathcal{L}_{\chi}+\mathcal{L}_{S} - g_{\varphi}\,S\,\bar{\chi}\,\chi + \mathcal{L}_{I} +\call_V\\
\mathcal{L}_{\chi}&=&\bar{\chi}\,(i\slashed{\partial}-m_{\chi_0})\chi \\
\mathcal{L}_S&=&\frac{1}{2}\partial_{\mu}S\partial^{\mu}S- V_S\\
\mathcal{L}_{I}&=&-\lambda_1\Phi^{\dagger}\Phi S-\lambda_2\Phi^{\dagger}\Phi S^2
\\
\call_V&=&-\frac{1}{4}V_{\mu\nu}V^{\mu\nu}+\frac{1}{4}\lambda_V\,S^2\,V_{\mu}V^{\mu}-g_V\,\bar{\chi}\,\slashed{V}\,\chi 
\eea
where $ \call_{SM} $ is the Lagrangian density for the  SM; $\Phi$ the SM Higgs and $S$ a scalar field connecting with the dark sector;
$\chi$ is the dark fermion singlet; the potential in $\call_S$ is defined as
$V_S=\frac{m_0^2}{2}\,S^2+\frac{\lambda_3}{3!}S^3+\frac{\lambda_4}{4!}S^4$.
The dark vector boson part is  $\call_V$, with the field tensor given by $V_{\mu\nu}=\partial_{\mu}V_{\nu}-\partial_{\nu}V_{\mu} $.
The form of  $V_S$ clearly indicates the possibility of spontaneous broken
symmetry in the model, which is parameterized by
introducing $ \langle S \rangle = x_0 $. Considering  spontaneous symmetry breaking in both sectors, the writing field SM
Higgs as 
$\Phi=\frac{1}{\sqrt{2}}\left( \begin{array}{c}
0   \\
v_0+h(x)
\end{array} \right)$
and $ S = x_0 + \varphi $, where $ \call $ contains a quadratic form in $h$ and $ \varphi $, indicating that these are not mass eigenstates theory.

There remains the possibility that the interaction between the SM and the dark sector only  through the scalar field: $S$ with 
SM occurs exclusively
through the Higgs boson, which includes the model in a class of theories known as {\it Higgs portal dark matter}. 
Writting $\calv=V_S+V(\Phi^{\dagger}\Phi)$ and
demanding that $  \left.\frac{\partial \calv}{\partial h}\right|_{v_0,x_0}= \left.\frac{\partial \calv}{\partial S}\right|_{v_0,x_0}=0$,
one obtains
\bea
m_0^2=-\frac{\lambda_3}{2}x_0-\frac{\lambda_4}{6}x_0^2-\lambda_2v_0^2-\frac{\lambda_1v_0^2}{2x_0}
\hs
\mu^2=\lambda v_0^2+x_0(\lambda_1+\lambda_2 x_0)\,.
\eea
The neutral scalar states $h$ and $\varphi$ defined by $H_0 = (v_0 + h)/√2$ and $S = x_0 + \varphi$ are mixed
to yield the mass matrix given by \cite{singlet}
\bea
M^2_{hh}&=& \left.\frac{\partial^2 \calv}{\partial h^2}\right|_{h=\varphi=0} =2\lambda v_0^2 
\hs
M^2_{ss}=\left.\frac{\partial^2 \calv}{\partial \varphi^2}\right|_{h=\varphi=0} =\frac{\lambda_3x_0}{2}+\frac{\lambda_4}{3}x_0^2-\frac{\lambda_1v_0^2}{2x_0} 
\nn\\
M^2_{hs}&=&M^2_{sh}=   \left.\frac{\partial^2 \calv}{\partial \varphi\partial h}\right|_{h=\varphi=0} =     (\lambda_1+2\lambda_2x_0)v_0\,.
\eea 
The two physical Higgs $h_1$ and $h_2$ are  mixtures and the mass eigenstates are obtained by
\bea
h_1&=&\varphi\, \sin\theta+h\, \cos\theta \nn\\
h_2&=&\varphi\, \cos\theta-h \,\sin\theta
\label{higgs}
\eea 
the mixing angle $\theta$ is defined by
\bea 
\tan\theta=\frac{y}{1+\sqrt{1+y^2}}
\hs
y&=&\frac{2M^2_{hs}}{M^2_{hh}-M^2_{ss}}\,.
\eea 
The $h_1$ and $h_2$   masses are
\bea 
m_{1,2}^2=\frac{M^2_{hh}+M^2_{ss}}{2}\pm\frac{M^2_{hh}-M^2_{ss}}{2}\sqrt{1+y^2}\,.
\eea 
In the present, the SM Higgs will be considered the $h_1$ field with mass $m_1=125$ GeV.
 
 \section{INTERACTION OF DARK PHOTONS}
 
\begin{figure}[t]
  \epsfig{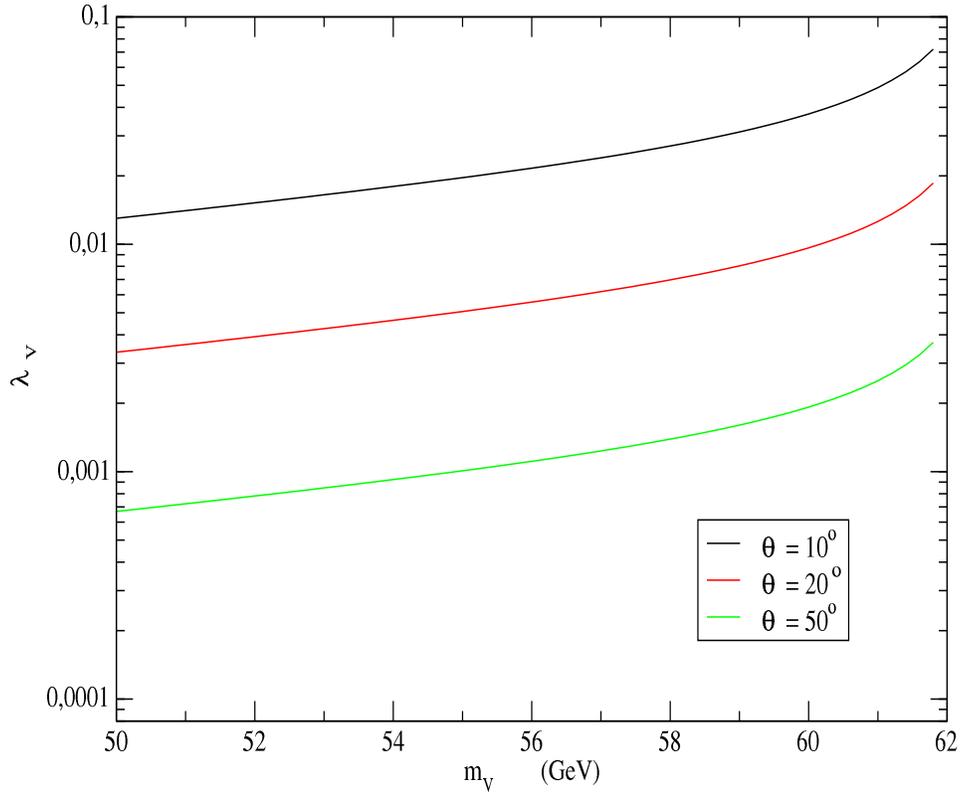}
  \caption{Parameter space relating the coupling $\lam_V$ to the dark photon mass $m_V$. }
  \label{lamv-fig}
\end{figure}

 The connection of the SM with the dark sector is through the interaction with the scalar field $S$. In particular
 the term that describes the interaction between vector $V_\mu$ (dark photon) and $S$ is $\frac{1}{4}\lambda_V \,S^2\,V_{\mu} V^ {\mu} $. 
 After spontaneous symmetry breaking this term changes as
 \bea 
 \frac{1}{4}\lambda_V \,S^2\,V_{\mu} V^ {\mu}\to \frac{\lambda_V\,x_0^2}{4}\,V_{\mu} V^ {\mu}
+\frac{\lambda_V\,x_0}{2}\,\varphi\,V_{\mu} V^ {\mu}+\frac{\lambda_V}{4}\,\varphi^2\,V_{\mu} V^ {\mu}\,.
 \label{ssb}
 \eea 
 The first term on the RHS of (\ref{ssb}) clearly reveals a mass term $m_V^2/2$ for the dark photon. 
 This term shows  that the  coupling $\lambda_V$ is related to the scale $x_0$ and to the vector boson's mass $m_V$ by  
\bea 
\lambda_V=2\,\frac{m^2_V}{x_0^2}\,.
\label{vmass}
\eea  
 By means of    (\ref{vmass}) one writes the two interaction Lagrangians of $V_\mu$ with $ \varphi $ as
 \bea
 \call_{\varphi\,V} = \sqrt{\frac{\lambda_V}{2 }}\,m_V \,\varphi\, V_{\mu} V^{\mu}
\hs
\call_{\varphi\varphi\,VV} =\frac{\lambda_V}{4}\,\varphi^2\,V_{\mu} V^ {\mu}\,.
 \label{phi-v}
 \eea
In this model, to describe the decay of the physical Higgs field $h_1$  into dark photons, one 
substitutes  the $\sin\theta\,\, h_1$ contribution in $ \varphi $ of (\ref{higgs}) into $\call_{\varphi\,V}$ 
 obtaining a new expression:
  \bea
\call_{h\,V}=\sqrt{\frac{\lambda_V}{2 }}\,m_V\,\sin\theta\,\, h_1\,V_{\mu}V^{\mu}\,.
\label{hvv}
\eea 
From (\ref{hvv}) the Higgs partial decay width into a massive dark photon can be calculated resulting in
\bea
\Gamma\, (h_1\to VV)=
\frac{\lambda_V \,m_{1}^3\,\sin^2\theta}{64\,\pi\, m_V^2}\left(1-4\frac{m_V^2}{m_{1}^2}+12\frac{m_V^4}{m_{1}^4}\right)\sqrt{1-4\frac{m_V^2}{m_{1}^2}}\,.
\label{decay}
\eea
The parameter space can be generated from (\ref{decay}) for  different mixing angles, assuming that this channel has a 
branching ratio $BR=\Gamma/\Gamma_{tot}$ of 10\%. We shall consider $\Gamma_{tot}=6.1$ MeV \cite{prl}. The result of this 
calculation is seen in Fig. (\ref{lamv-fig}).

\begin{figure}[h]
  \epsfig{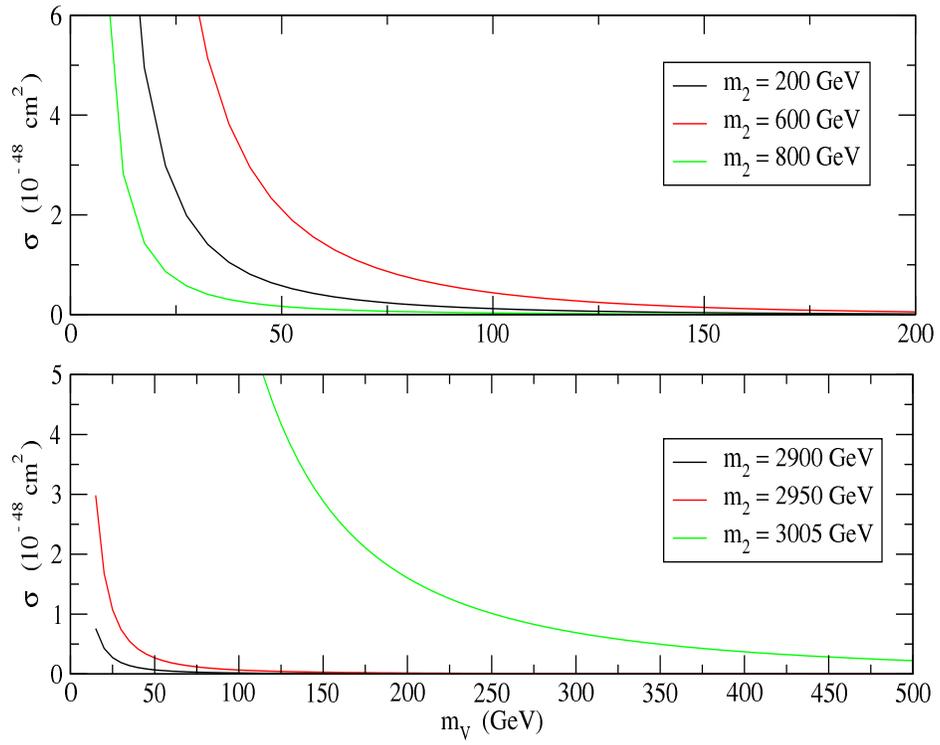}
  \caption{ Mixing angle $10^o$, $\lam_V \sim 1$,  $\sqrt{s}=$ 500 GeV (upper) and $\sqrt{s}=$ 3 TeV (lower).}
 \label{eevv}
\end{figure}

The chances of observing dark photon pair production in the continuum might be higher in a cleaner environment 
like in $e^{-}e^{+}$ collisions. 
 The  $e^{-}e^{+}\to VV$ annihilation cross-section is
 \bea 
 \sigma= \frac{\lambda_V}{4\pi}
 \sqrt{1-4\,\frac{m_V^2}{s}}
 \left(\frac{m_e\,m_V\,(m_2^2-m_1^2)\,\sin(2\,\theta)}{4\, v_0}\right)^2
 \left(2+\frac{(s-2m_V^2)^2}{4m_V^4} \right)
 \frac{1}{ (s-m_1^2)^2\,(s-m_2^2)^2}
 \label{cross}
 \eea 
We have evaluated this cross-section at  the energy range relevant for the ILC, $\sqrt{s}$ = 500 GeV
and at $\sqrt{s}$ = 3 TeV relevant for the CERN CLIC \cite{plb}. The other  parameters used are
mixing angle $10^o$ and the coupling $\lam_V \sim 1$.  A comparison for different masses $m_2$ of the scalar field
is considered. At this low value for the angle $\theta$ the mixing is small such that  $h_1\sim h$ and $h_2\sim \varphi$.
The results are shown in Fig. (\ref{eevv}).
 Current efforts of studying Higgs physics at the LHC   may  result in difficulties to observe the dark matter effects. 
 A solution could be to perform  these studies  of Higgs decays at   $e^{-}e^{+}$ colliders. 
The fundamental drawback is that  dark matter particles 
have extremely small pair production cross sections.  As shown in the example of $e^{-}e^{+}\to VV$, they are too small to be observed  in colliders 
unless an enhancement for 
very high luminosities are made available.
 

\begin{figure}[h]
  \epsfig{file=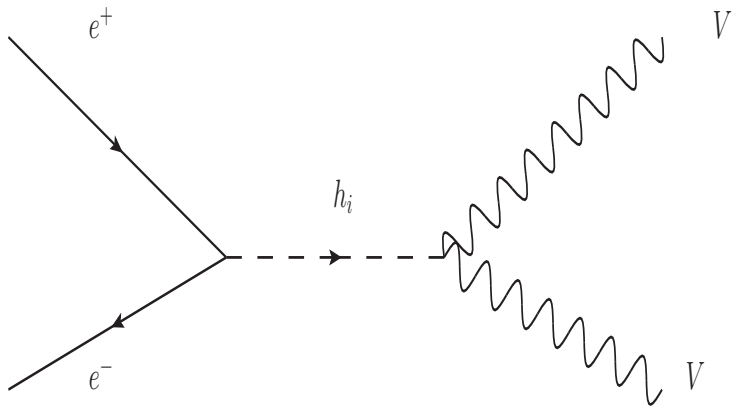, height=170.00pt, width=300.00pt,angle=0}
  \caption{$VV$ production from $e^{-}e^{+}$ annihilation.}
  \label{higgs-fig}
\end{figure}

\section{ACKNOWLEDGMENTS}

This research was supported by Conselho Nacional de Desenvolvimento 
Cient\'{\i}fico e Tecnol\'{o}gico (CNPq) and Universidade Federal do Rio
Grande do Sul (UFRGS).


\end{document}